\documentclass[%
 reprint,
 amsmath,amssymb,
 aps,
pre,
]{revtex4-2}

\usepackage{graphicx}
\usepackage{dcolumn}
\usepackage[utf8]{inputenc}
\usepackage{newunicodechar}
\newunicodechar{̈}{\"{}}
\usepackage{bm}
\usepackage{xcolor}
\usepackage{hyperref}
\graphicspath{ {./Figures/} }

\begin{document}

\preprint{APS/123-QED}

\title{Geomimicry: Emergent Dynamics in Earth-Mediated Complex Materials}

\author{Shravan Pradeep$^{1,2}$, Emanuela Del Gado$^{3}$, Douglas J. Jerolmack$^{1,2}$, and Paulo E. Arratia$^{2}$}
\email{Corresponding author: parratia@seas.upenn.edu}

\affiliation{
 $^{1}$Department of Earth and Environmental Science, University of Pennsylvania, Philadelphia, PA 19104, USA\\
 $^{2}$Department of Mechanical Engineering and Applied Mechanics, University of Pennsylvania, Philadelphia, PA 19104, USA\\
 $^{3}$Institute for Soft Matter Synthesis and Metrology, Georgetown University, Washington D.C. 20057 , USA
}

 \date{\today}

\begin{abstract}
    Soils and sediments are soft, amorphous materials with complex microstructures and mechanical properties, that are also building blocks for industrial materials such as concrete. These Earth-mediated materials evolve under prolonged environmental pressures like mechanical stress, chemical gradients, and biological activity. Here, we introduce \textit{geomimicry}, a new paradigm for designing sustainable materials by learning from the emergent and adaptive dynamics of Earth-mediated matter. Drawing a parallel to biomimicry, we posit that these geomaterials follow evolutionary design rules, adapting their structure and function in response to persistent natural forces through locally evolved interactions and compositions. Our central argument is that by decoding these rules--primarily through understanding the emergence of novel exotic properties from multiscale interactions between heterogenous components--we can engineer a new class of adaptive, sustainable matter. We propose two complementary approaches here. The top-down approach looks to nature to identify building blocks and map them to functional groups defined by their mechanical (rather than chemical) behaviors, and then examine how environmental training tunes interactions among these groups. The bottom up approach seeks to leverage and test this framework, building earth materials one component at a time under fluctuating environmental stresses that guide assembly of complex and out-of-equilibrium materials. The goal is to create materials with programmed functionalities, such as erosion resistance or self-healing capabilities. Geomimicry offers a pathway to truly design Earth-mediated circular materials, with potential applications ranging from climate-resilient soils and smart agriculture to new insights into planetary terraforming, fundamentally shifting the focus from static compositions to dynamic, evolving systems that are mediated via their environment.
\end{abstract}

\maketitle

\section*{Preamble}

Over the 50+ years since Philip Anderson’s 1972 landmark work, \textit{More is Different} \cite{anderson1972more}, the study of emergent dynamics in hierarchical systems has become central to condensed matter physics. The fundamental idea of our work is to leverage this framework--previously applied in biological matter \cite{artime2022origin,couzin2005effective}, quantum systems \cite{yennie1987integral,kivelson2016defining}, turbulent fluids \cite{lorenz1963deterministic}, and materials \cite{tong2020emergent} --to deepen our understanding of the assembly, dynamics, and transport in Earth-mediated matter. We define Earth-mediated matter as materials that have evolved over geologic timescales, developing unique mechanical, chemical, and transport properties under nature's stressors and through its architected design. Their microstructure governs important \emph{static} (e.g. carbon sequestration, groundwater storage, nutrient delivery etc.) and \emph{dynamic} (e.g. erosion, infrastructure stability etc.) functions of the near and sub-surface soil. Thus, here we focus on natural materials that are physico-chemically transformed under environmental stresses and are persistent under the Earth's climatic conditions today. This definition, while seemingly straightforward, would undoubtedly carry different implications for engineers and scientists, a divergence stemming from their distinct training and approaches. 

\begin{figure*}[t]
\centering
\includegraphics[scale = 0.65]{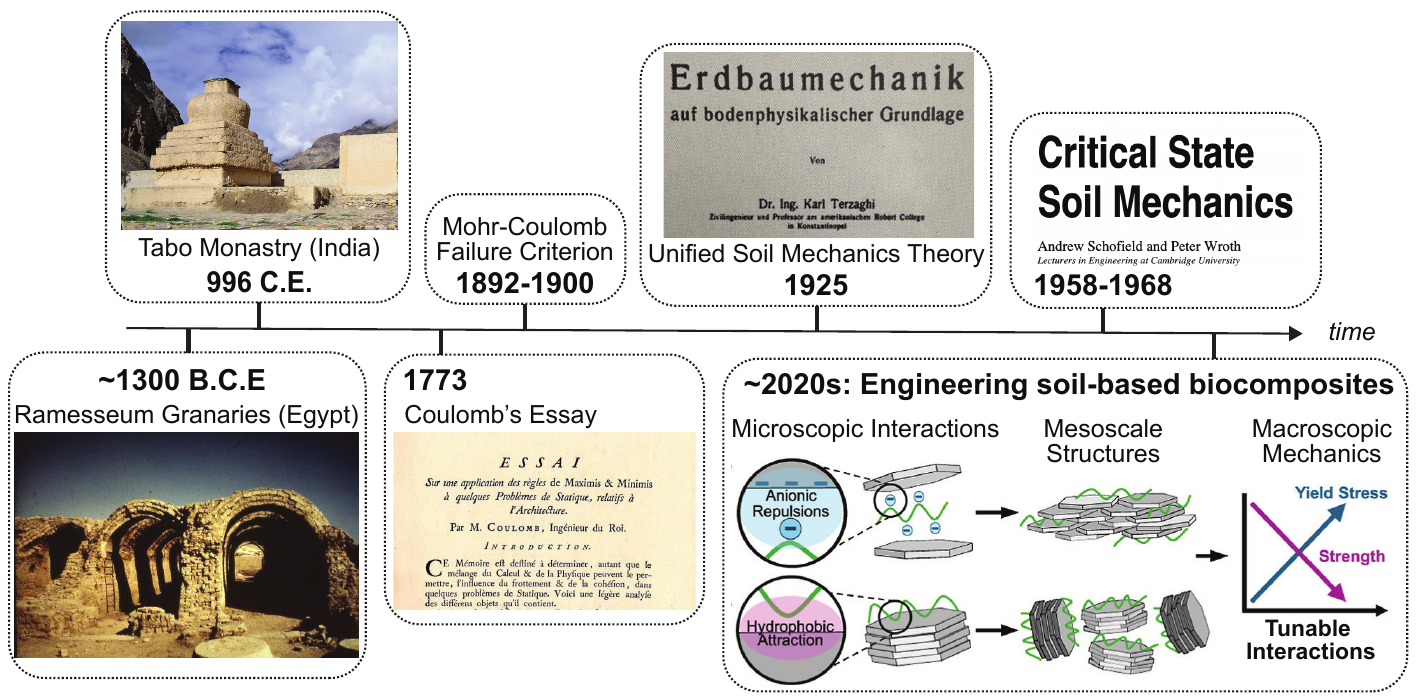}
\caption{\textbf{Timeline of soft earth engineering development: from ancient earth constructions to modern engineered composites.} The oldest standing adobe structures in the world are the granaries at the Ramesseum, built around 1300 B.C. by Ramses II near Luxor, Egypt. Adobe construction techniques spread across civilizations; one notable example is the Tabo Buddhist Monastery in the Spiti Valley, India, constructed around 996 C.E. In 1776, the French engineer and physicist Charles-Augustin de Coulomb published his now-famous essay \textit{Essai sur une application des règles de Maximis et de Minimis à quelques Problèmes de Statique relatifs à l’Architecture}, discussing soil shear strength. Building on these concepts, Christian Otto Mohr conceptualized the Mohr Circle in 1882 and later published it in 1900. The Mohr–Coulomb failure criterion relates shear stress ($\tau$), normal stress ($\sigma$), and cohesion ($c$) through the friction angle ($\alpha$) as: $\tau=c+\sigma$tan($\alpha$). Twenty-five years later, Carl Terzaghi’s \textit{Erdbaumechanik} (1925) launched modern soil mechanics, incorporating concepts from geophysics, physics, and mechanics, and establishing the fundamental concept of effective stress—thereby distinguishing geotechnics from other branches of engineering mechanics. A foundational paper in 1958 \cite{roscoe1958yielding} and the textbook that followed in 1968 \cite{schofield1968critical} provided a unified constitutive model explaining the complete mechanical behavior of soil, from initial loading to ultimate failure, based on the principles of plasticity theory and effective stress. Today, these foundational ideas have evolved to support the design of adaptive and resilient materials by engineering multiscale soil composites, where microscopic interactions are tuned to achieve desired macroscopic mechanical properties \cite{mikofsky2025physicochemical}. Image Courtesy: Pixabay; Wikimedia; and Development Workshop Digital Archive under Creative Commons Attribution Non-Commercial. 
}
\label{fig1}
\end{figure*}

From a fundamental perspective,  one may be drawn to focus on elucidating the underlying physical and chemical principles governing the formation and behavior of these natural materials, seeking to build comprehensive theoretical models. Despite its ubiquity and importance, however, Soft Condensed Matter Physics has not examined the assembly and dynamics of soil with the curiosity and rigor of a cell or a polymer gel. From a practical perspective, one may be primarily interested in the applications of Earth-mediated materials, aiming to design and optimize systems that harness their unique properties for specific functions. Certainly some corners of engineering, especially Geotechnics, have been developing such applications with soil; however, contributions from other relevant engineering disciplines have thus far been limited, despite the centrality of concepts from these fields. This perspective is not intended as an authoritative review that summarizes previous work. Instead, it is meant to entice scientists from Soft Condensed Matter (broadly defined) to begin work on understanding earth materials as hierarchical, far-from-equilibrium systems whose complex mechanical behaviors emerge from novel combinations of materials and stresses. It is also meant to encourage researchers to draw inspiration from earth materials; that is, to develop a geomimicry approach for sustainable materials comparable to efforts in biomimicry. By bridging the gap between abstract scientific arguments and practical  outcomes, it is our hope that this perspective helps to open up new and compelling questions for future research at the intersection of soft matter physics, earth sciences, materials science, transport phenomena, and engineering mechanics. 

\section*{Introduction} 

Humans (and animals) have developed a working knowledge of soil and earth materials over millennia (Fig. \ref{fig1}). This knowledge spans from early uses of adobe and rammed earth in ancient structures like the pyramids ($\sim$ 3,000 B.C.) to Coulomb’s 1776 pioneering work on friction and cohesion in earth materials, which laid the theoretical foundation for Geotechnical Engineering later \cite{coulomb1973essai,gillmor2017coulomb}. Traditional empirical and theoretical approaches merged into the framework of Critical State Soil Mechanics (CSSM), which integrates Mohr-Coulomb failure criteria with granular plasticity that accounts for dilation and compaction \cite{schofield1968critical,roscoe1958yielding}. This framework is central in Geotechnical Engineering, and has seen successful application in engineering design; for example, modern rammed earth buildings use carefully tuned heterogeneous particle mixtures that maximize compressive strength of frictional grains and tensile strength of cohesive particles \cite{BatirEnTerre}. Building on this foundation, sustainability and resilience drive innovations like self-healing soils and green cement \cite{dejong2010bio,phair2006green,hamza2020effect,Habert2020environmental,zaidi2021use}. Recent advances leverage soft matter, such as polymers and living matter, in designing sustainable built environments by expanding mechanical behavior, blending ancient working knowledge with modern engineering in composite materials (Fig.\ref{fig1}) \cite{fatehi2021biopolymers,chang2016introduction,mikofsky2025physicochemical,armistead2023toward,biggerstaff2022determining,lepech2025composite,lerma2024new}.

Recent studies in soft matter have begun to move beyond idealized materials, like colloidal suspensions and gels or dry granular materials, and examine the behavior of more complex mixtures that approach relevance for earth materials. The simplest starting framework is to assume that distinct components in these mixtures each confer a resisting stress, and that these stresses are additive. In other words, the components do not interact in a complex way. This approach has found success in describing the rheologic behavior of some systems, for example: moist sand \cite{herminghaus2005dynamics,louati2015apparent}, emulsions \cite{mason1996yielding}, and granular particles in non-Newtonian carrier fluids \cite{khruslov2004homogenized,kammer2022homogenization,chateau2008homogenization,ovarlez2015flows}. Such additive properties are exceptions in multicomponent soft materials, not the norm. In most cases, novel behaviors \emph{emerge}, where the whole differs from the sum of its parts. This is particularly evident in soft matter systems, where complex interactions give rise to unexpected macroscopic properties. For instance, granulation--whether in particulate mixing, soil wetting, or chocolate conching--reveals dynamics more intricate than classical models suggest, governed by fragile dynamics, amorphous microstructures, and competing mesoscopic length scales \cite{blanco2019conching,hodgson2022granulation,jerolmack2019viewing,voigtlander2024soft}. Recent interest in colloidal gel-based composites demonstrate how introducing secondary phases (e.g. nanoparticles, granular inclusions, or multiple networks) has dramatic impact: from reinforcement \cite{Dellatolas2023local} to shifting gel percolation thresholds through arrested gelation \cite{jiang2025filled,jiang2023colloidal}, from creating flow-induced strengthening and novel bi-stable states \cite{jiang2022flow,li2023impact} to complex architectures \cite{MugnaiPNAS2025}. These systems exhibit emergent properties when particles interact with polymers or gels, leading to mesoscale structuring that defies additive predictions using the properties of their individual phases. Our recent work examining model soft Earth suspensions made of frictional, cohesive and viscous elements confirms this idea, where we identified new controls of material properties on universal behaviors and critical points \cite{pradeep2024origins,kostynick2022rheology}.

Here, we seek to (re-)introduce the term ``Geomimicry'', as a scientific framework and materials design approach. Geomimicry examines Earth-mediated materials as novel states of matter, hierarchical, multi-component materials assembled by broad-band transient stresses, and draws inspiration from their exotic mechanical properties and resilience to environmental stresses, to build novel sustainable materials. Although the terms ``geomimicry'' and ``geomimetics'' have been used in the past \cite{williams2017geomimicry,gilbert2022geomimetism,dumas2016fast,kim2022geomimetic,goure2014interfacial,goure2014understanding,lerma2023novel,yang2015organic,mao2022geomimicry,massion2022geomimicry,bockisch2019selective,meunier2020geoinspired}, our re-definition seeks to align the term and its aims with the more popular Biomimicry. Forms in nature are examined in biomimicry as optimized structures shaped by the selective pressure of the environment, and design principles are drawn from them. We hope this perspective provides the foundation for scientists and engineers to draw similar inspiration from Earth-mediated materials. 

\section*{Biomimicry \& Geomimicry }

\begin{figure*}
\centering
\includegraphics[scale = 0.75]{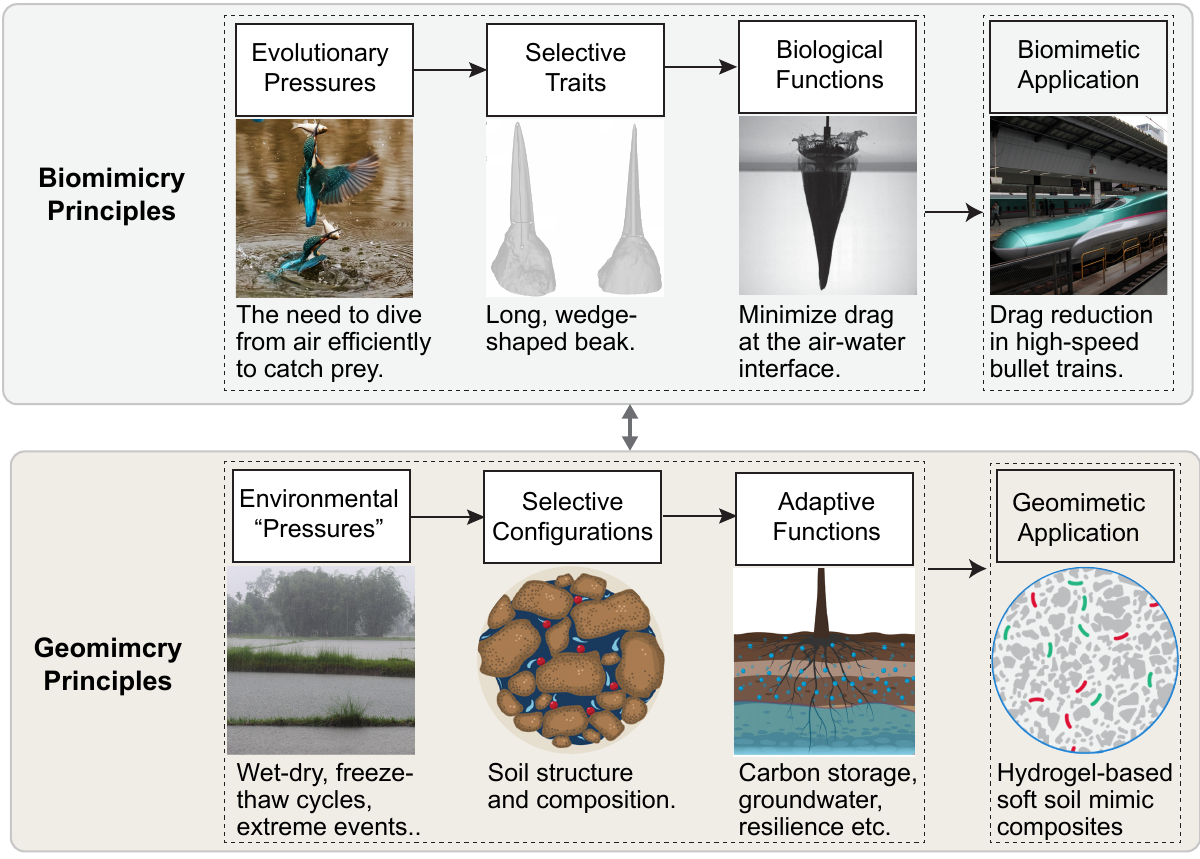}
\caption{\textbf{Drawing parallels between biomimicry and the proposed geomimicry framework}. The biomimicry framework focuses on understanding the final biological functions that emerge through selective traits evolved under external pressures, and applying these insights to engineering design. The example illustrated shows how the need to catch prey from water led to the evolution of the kingfisher’s long, wedge-shaped beak, which minimizes drag at the air–water interface, a principle later applied to reduce drag in high-speed bullet trains. Similarly, the proposed geomimicry framework mirrors this concept: soil functions arise from evolved configurations of soil microstructure, shaped by real environmental pressures. This framework can be used to design multicomponent, multifunctional soil-mimetic composites. Image courtesy: Pixabay; images adapted and modified from literature \cite{hochman2021diverse,white2020water,smercina2022synthetic}.}
\label{fig2}
\end{figure*}

Over the past 3.8 billion years since life first emerged on Earth, it has continuously evolved toward more complex forms that are adapted to their environment. The biological functions of life forms on Earth are a result of evolutionary pressures, which act as an energy input or engine that drives adaptation to the environment \cite{sadava2011life,kauffman2019world,judson2017energy}. These pressures define the objective functions and constraints; biology experiments by random mutation, and robust ``traits" in organisms emerge when mutations serve functions that are useful for survival. For example, high-contrast pelage patterning on Zebras serves to reduce fly landings \cite{caro2014function,caro2023don}, while the Kingfisher's elongated beak profile aids in drag reduction at the air–water interface for catching prey (Fig. \ref{fig2}) \cite{crandell2019repeated,eliason2020morphological}. Evolutionary pressure has pushed many extant species to develop micro- or nano-structures that can significantly affect material functionalities, such as aerodynamics \cite{sirohi2013engineered,billingsley2021biomimicry,kozlov2015bio,primrose2020biomimetics,kaya2025aerodynamic}, wettability \cite{liu2012extreme,solga2007dream,xu2016biomimetic,bhushan2009biomimetics}, and optical properties \cite{dumanli2016recent,crne2011biomimicry,tadepalli2017bio,wright2013engineered}, to name a few. Biomimicry is the framework that learns from the relations between biological evolutionary pressures and optimized traits, and uses them to engineer novel materials. A key principle is identifying relevant operating windows and length scales, and harvesting design principles, such as structural hierarchy \cite{pan2014exploring,chen2015lessons,du2019looking}, gradients \cite{qu2015engineering,passino2002biomimicry,dong2020progress}, textures \cite{jullien2020biomimicry,ivanovic2018biomimetics,eadie2011biomimicry}, and mechanical properties \cite{gebeshuber2011biomimetics,dicks2016philosophy,grigorian2014biomimicry}. A famous early example of biomimicry is Leonardo da Vinci's study of flight, in which he analyzed the wing motion of birds/bats to design a flexible, jointed wings' device known as the Ornithopter in the late 15th century \cite{vogel2013comparative,gerdes2012review}.  

Earth-mediated materials, such as soil and soft sediments, are akin to living matter. They have been on the surface of Earth for even longer than life ($\sim$4.5 billion years), and have evolved increasingly complex composition and structure over geologic time. Indeed, soils and sediments on Earth have co-evolved with life; examples include the microbe-facilitated precipitation of carbonate (including coral reefs) in the ocean \cite{reid2000role_Nature}, and the development of organic-rich soils on land that followed the emergence of terrestrial plants \cite{beraldi2013early,lenton2016PNAS}.  To develop the geomimicry framework, we attempt to map the biological concepts of ``evolutionary pressures'' and ``traits'' to describe geological material evolution. Earth-mediated materials are subject to a barrage of environmental stresses in the form of: cycles of humidity, temperature and atmospheric pressure; fluid stresses ranging from capillarity and buoyancy effects to shearing by currents of water and wind; and shaking by earthquakes, just to name a few. Biotic factors such as microorganisms (fungi, bacteria, protists, and archaea), worms, animals, human activity, and plant roots also act as mechanical disturbances.

The evolutionary pressures we consider are thus actual ``mechanical stresses". The effects of these external mechanical stresses manifest at the scale of the constitutive particles--in terms of sizes, shapes, and interparticle interactions, micro- and meso-scale structures formed, and the overall locally evolved composition. Often the evolution of Earth-mediated materials like soil is gradual, changing composition over millennia in response to innumerable cycles of (mechanical) stresses \cite{lin2011three}. Intermittent events such as wildfires, however, can dramatically alter interparticle interactions, thus modifying soil chemistry and its resultant mechanical properties instantaneously. This makes soil and soft sediments a novel class of \emph{adaptive matter}, where their microstructure and composition is a direct consequence of the spectrum of environmental loads under which they were evolved. In other words, Earth-mediated materials can sustain self-stress because their interparticle interactions have been shaped by the prevailing microclimate through prolonged environmental forcing. For instance, wet–dry cycling drives capillary condensation of fine particles into solid bridges among larger grains results in emergence of a cohesive function from the specific forcing history rather than convergence toward a unique optimum \cite{seiphoori2020formation}. The composition, pore architecture, and grain-scale properties of Earth-mediated materials preserve the imprint of the geologic processes under which they have evolved; this means that the geologic record is an archive of extreme events and past stresses -- if we can learn to decode it \cite{sadler2015scaling,jerolmack2010shredding}. 

While the analogy between biomimicry and geomimicry is productive, an important distinction must be acknowledged. Biological systems are active: organisms sense environmental cues, generate metabolic energy, and exhibit adaptive variation that is heritable. The geophysical systems we consider here lack agency. They are passively shaped by external forcing. Yet even within this scope, Earth-mediated materials experience environmental noise, that drive these materials out of equilibrium and can lead to spontaneous reorganization (aging), threshold-driven responses (yielding and cracking), and history-dependent behavior (creep and memory). Thus, both biological and Earth-mediated materials can be viewed through the lenses of non-equilibrium systems, with appropriate distinctions. 
For e.g., geomaterials are responsive without being autonomous. The resulting design principles may be more directly transferable to engineering than those extracted through biomimicry since the optimization in Earth-mediated materials is governed by mechanics and thermodynamics rather than by genetics and ecology. We note that ``evolution" is used throughout this perspective in the physicist's sense, temporal change of a system's state under external driving, rather than the biologist's sense of descent with modification under natural selection.

By mapping evolutionary pressures and traits from biomimicry to mechanical stresses and soil composition, respectively, we define the research area of ``geomimicry" as a framework that identifies the evolution-composition relationships in Earth-mediated matter. Mirroring biomimicry, one can pursue complementary strategies. A \emph{top-down approach} starts from observing nature and its processes to determine the categories of constituent materials, interparticle interactions, environmental stresses, and the associated dynamics. In contrast, a \emph{bottom-up approach} focuses on designing earth materials one component at a time -- optimizing particle size, shape, polydispersity, interactions, and additives -- and ``training" the mixture with varied stress regimes to engineer relevant mechanical properties such as strength, toughness, ductility, and resilience. In both cases, mesoscopic length and time scales emerge from materials' microscopic constituents, interactions, and their training. Here, we propose a framework for analyzing geomaterials through an \textbf{evolutionary lens}, structured around the following key elements: 
\begin{itemize}
    \item \textbf{Stressors}: Earth materials are exposed to a wide range of local environmental or mechanical stresses. Thus, one needs to examine how those stresses and boundary conditions affect material properties and mesoscale structure.  
    \item  \textbf{Adaptation}: External constraints progressively train the constituents of Earth materials, shaping their interactions to produce mesoscale structures and unique properties under local climatic conditions. It is important to establish quantitative relationships linking the material’s mesoscale architecture (e.g., network structures) to its macroscopic properties (e.g., fracture toughness). 
    \item \textbf{Functionality}: Identify and map how new functionalities arise from the interplay between external stresses and microstructural evolution. 
\end{itemize}
Following these steps will help us understand the \textit{adaptive} nature of soft sustainable geomaterials.

\section*{Emergence in Earth-Mediated Matter}
The guiding principle of emergence in many-body physics problems was well articulated by Anderson: ``We expect to encounter fascinating and... very fundamental questions at each stage in fitting together less complicated pieces into the more complicated system and understanding the basically new types of behavior which can result'' \cite{anderson1972more}. In soft matter and fluid physics, it is common to observe the emergence of dissipative, mesoscopic structures (between particle and system scale) whose behavior governs the bulk dynamics; eddies in turbulence \cite{kim2024early}, force chains in granular media \cite{Juanes2024dynamic,nampoothiri2020emergent}, and quadrupolar (Eschelby) strain fields in a broad range of amorphous solids \cite{falk1998dynamics,falk2024topological,Maloney2006Amorphous,deshpande2021perpetual}. Emergent structures make these systems more `complex', in that these mesoscopic scales usually cannot be predicted from the fundamental particle scale. But emergent scales are also key for \textit{universality}. That is, robust and generic properties shared by different systems that are insensitive to microscopic details, and amenable to theoretical approach (e.g., statistical mechanics). 

Emergence has been recognized in geoscience as well, most notably in the context of understanding landscape patterns across scales \cite{werner1999complexity}. Entrainment of a single sand grain from a river bottom remains a formidable prediction challenge, due to the complexities of turbulent wall stresses along rough boundaries and also granular interactions in heterogeneous beds \cite{schmeeckle2001interparticle,schmeeckle2003direct}. Yet, innumerable particles organize into sand dunes, whose emergent dynamics are controlled by fluid-structure interactions at the scale of river depth, that do not require detailed knowledge of grain-scale physics. Moving up another order of magnitude in scale, rivers form meander bends due to fluid-structure interactions at the scale of river width; dunes only enter into this problem as an averaged bed friction. At the largest scales, river networks are famously fractal \cite{rodriguez1994self,rodriguez1997fractal}, and can be understood using generic frameworks akin to diffusion-limited aggregation \cite{devauchelle2012ramification,devauchelle2017laplacian}. Laboratory-scale landscape experiments may reproduce such universal patterns when systems exhibit internal similarity, wherein the interplay of structural and geomorphic self-organization generates scale independence \cite{paola2009unreasonable}. Collectively, these examples suggest that universality in geological systems may emerge from interaction networks hierarchically organized rather than simply from the microscopic details of their constituents. At the same time, the robustness observed in natural systems implies that their macroscopic behavior is more plausibly governed by the emergent physics of collective interactions than by the characteristics of individual components \cite{Gao2003materials}. In essence, we seek universal mechanisms that operate across different scales that lead to a desired material functionality that is more or less independent of constituent-level details. Building on this view, the present Perspective will consider how material properties mediate and constrain macroscopic landscape dynamics.

To understand the evolution of microstructure and the associated mechanical properties of complex earth materials, it is necessary to move beyond simple model systems and the additive treatment of material functions. Here, we introduce a set of terminologies that will anchor the central concepts of materials geomimicry. Earth-mediated matter are soft composite mixtures comprising grain-scale particulates (sand, clay, and silt), soft geopolymers and extracellular polymeric substances, microfauna, and pore spaces that are partially or fully saturated with water. These components collectively form the \emph{building blocks} of soft Earth matter. The central hypothesis of geomimicry is that each of these microscale building blocks possess \emph{mechanical functional groups} that undergo \emph{environmental training} producing soil compositions (the geological ``traits") that manifest unique \textit{macroscopic properties}, each reflecting a distinct evolutionary trajectory. In what follows, we elaborate on how macroscopic properties emerge from these microscale building blocks. 

\subsection*{(Non)Ideal Behavior in Ideal Systems}

We begin by reflecting on the limitations and challenges of applying model system frameworks from soft matter physics, particularly oscillatory and granular rheology, as these represent the laboratory-scale approaches most analogous to environmental pressures. Although model soft matter systems are often regarded as ideal platforms for rheological study due to their tunable mechanical properties, their responses can become profoundly non-ideal and geometry-dependent once they are driven beyond the linear regime. In oscillatory tests, increasing strain amplitude disrupts the underlying microstructure, giving rise to nonlinearities such as harmonic distortions \cite{liang2020distortion}, intracycle strain-stiffening or softening \cite{Ewoldt2008new,mermet2015laos,Donley2022Time}, as well as instabilities from yielding \cite{divoux2024ductile} to wall slip \cite{ballesta2012wall}, edge fracture \cite{chan2023perspective}, and adiabatic (viscous) heating \cite{giacomin2012viscous}. All of these features complicate the extraction of intrinsic material properties \cite{yang2017dynamic}. Likewise, in granular rheology, the assumption of a homogeneous, local response breaks down due to non-local effects; phenomena such as shear banding and creep are strongly influenced by geometric confinement, boundary conditions (wall roughness), and system size \cite{ovarlez2001rheology,schuhmacher2017wall}. Consequently, the measured response often reflects the particle–geometry system rather than the  material itself. These observations highlight a central paradox: while controlled laboratory systems reveal the fragility of “ideal” rheological frameworks, natural earth systems display robustness and reproducibility across scales. This contrast motivates the need for more general frameworks that can account for the emergent mechanical behavior observed in geomaterials.

\subsection*{Formation of Natural Geomaterials}
In developing general principles for complex natural systems, we first explore a \emph{top-down} approach within the framework of materials geomimicry.\\

\noindent
\textsl{Mechanical Functions.} In the geomimicry context, mechanical function denotes a behavioral category, e.g., cohesion, friction, squishiness (compliance arising from microscale compressibility), viscosity etc., rather than a specific composition. Materials that deliver the same response under an external load are grouped into the same mechanical functional group, directly analogous to chemical functional groups that classify atoms by characteristic reactivity. The same mechanical function can be achieved by different building blocks. Consequently, we treat distinct material chemistries as equivalent if they produce indistinguishable mechanical behavior at the relevant scales. For example, tuning the microscale surface roughness by manipulating microscale chemistry in experiments or using a conventional particle-scale friction coefficient in simulations has been shown to enhance macroscale mechanical properties, such as shear thickening and material elasticity \cite{minten2025hydrodynamic,pradeep2022hydrodynamic,pradeep2021jamming,hsu2018roughness,jamali2019alternative}. This indicates that the mechanical “frictional” functional group dominates the behavior, irrespective of its grain-scale origins. Formally, we can define a mechanical function as combination of intrinsic characteristics (e.g., surface energy, elasticity, swelling capacity, charge state, roughness) that yields a target constitutive response, independent of the origin of those characteristics. This means that convergent outcomes can arise from disparate microscopic interactions. 

\begin{figure*}
\centering
\includegraphics[scale = 0.60]{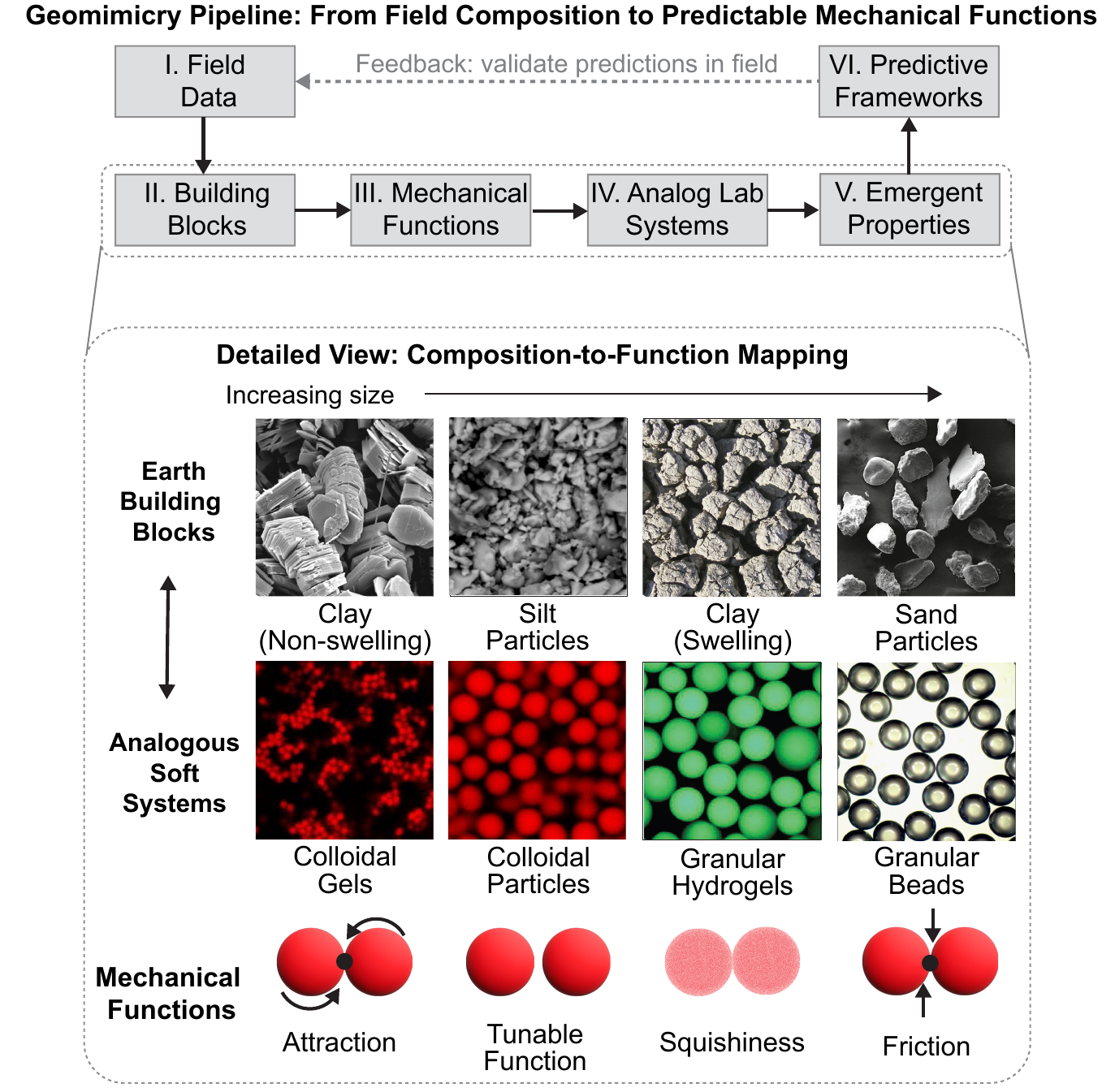}
\caption{\textbf{Geomimicry pipeline: from field composition to predictable mechanical functions.} The key steps are: (1) acquire field compositional data; (2) identify the constituent building blocks; (3) map each building block to its mechanical functional group(s); (4) construct analog laboratory systems that replicate the same functional combinations; (5) study emergent properties under controlled environmental training; and (6) use the resulting structure–property relationships to predict field-scale mechanical behavior. In a more fundamental sense, building blocks of Earth-mediated matter are mapped to their analogous soft particulate systems through their similar mechanical functional groups. Model soft particulate systems that mimic non-swelling clay (e.g., kaolinite), swelling clay (e.g., bentonite), and silica sand particles include colloidal gels (e.g., depletion-induced colloidal attraction), granular hydrogels, (e.g., carbopol), and granular beads, respectively \cite{qazi2022methods,wilson2014influence,zheng2015late,xiouras2018particle}. Silt particles and their analogous colloidal particulate systems exhibit complex interparticle potentials and are sensitive to small environmental changes, such as the presence of salts or polymers, resulting in tunable mechanical functions. The attraction and frictional mechanical functional groups are depicted as rolling and sliding constraints, respectively, as recently treated in suspension rheology community \cite{guy2018constraint,singh2020shear,singh2022stress}.}
\label{fig3}
\end{figure*}

Let us consider a few examples to illustrate this concept. Capillary bridges in moist sand \cite{Habert2020environmental} and sticky dry grains \cite{castellanos2005relationship} both increase tensile strength and yield stress of the respective material. Although the origins differ, liquid-mediated capillary forces on the former and interparticle adhesion on later, each implements a similar cohesive function at the continuum scale. Likewise, electrostatic attractions in suspensions of non-swelling clays such as kaolinite generate percolated networks with well-defined elastic moduli, which is functionally analogous to depletion-induced colloidal gels, even though the underlying chemistries differ. Swelling is another example, where there are separate routes to a distinct function: compliance or ``squishiness". Bentonite clays imbibe water and become grain-scale compressible, much like hydrogel beads. That microscale compressibility modifies bulk constitutive behavior in ways not available to incompressible hard particles such as sand or non-swelling clays. This makes ``squishiness" a distinct mechanical function, orthogonal to cohesion and friction; it explains why two clay-rich systems, kaolinite vs. bentonite, can share cohesive behavior yet different system compliance \cite{bourg2017clay}. 

Therefore, by organizing soft-earth materials around mechanical functions, such as cohesion, friction, squishiness, viscous dissipation etc., rather than composition, we gain a new material function-based framework for these complex systems.\\

\noindent
\textsl{Environmental Training.} While mechanical functions determine the (qualitative) behavior of soft Earth materials, their properties are further shaped by environmental training. Natural earth systems are continuously exposed to mechanical, chemical, and biological excitations arising from fluctuations in temperature, humidity, rainfall, fluid shear, earthquakes, wildfires, microbial mixing, and other disturbances. These drivers generate a wide spectrum of cycles with timescales spanning a couple of seconds to years \cite{Bowman2009FireEarthSystem,Bardgett2014BelowgroundBiodiversity}. Such cyclic forcing (or stresses) compels the fundamental building blocks of these Earth materials to collectively reorganize and interact across multiple length scales, gradually pruning unstable configurations until a metastable ``trait" emerges. This is a structure whose persistence timescale exceeds that of the dominant forcing, and that is in this sense locally adapted to the prevailing climatic conditions. Such configurations are not equilibrium configurations. They may continue to evolve under slower processes and can be pruned by new forcing protocols. This continuous programmability under environmental changes is a property that is central to natural geomaterials. 

At the simplest level, environmental training occurs under persistent unidirectional shear. This happens most clearly in river flows, where repeated stresses promote collisional sorting \cite{Claire2022threshold,jerolmack2007conditions, masteller2024fluvial} and the emergence of load-bearing clusters \cite{church2006bed,Frey2011BedloadGranular} and streamwise fabrics \cite{cunez2022strain} that resist entrainment. The onset of grain motion is classically governed by the Shields number, $\theta = \tau/(\Delta\rho gd)$, where $\tau$ is the shear stress, $\Delta\rho$ is the density difference between particles and the surrounding fluid, $g$ is the acceleration due to gravity, and $d$ is the average particle diameter. The parameter $\theta$ is thus the ratio of fluid shear stress to the submerged gravitational stress on a grain. Environmental training modifies $\theta$ itself, where the subcritical flows compact the bed and imprint anisotropic fabric, raising the critical Shields number for entrainment; a direct mechanical readout of stored memory. Analogous shear signatures are imprinted in fault gouge, where granular textures record cycles of slip \cite{houdoux2021micro,dorostkar2019grain}. On the next level, alternating wetting and drying provides a complementary chemo-mechanical training \cite{kleber2021dynamics}; capillary cementation during drying and swelling or softening upon rewetting drive the system far from equilibrium, producing aggregates \cite{seiphoori2020formation}, crusts \cite{routh2013drying, licsandru2023evaporative}, and desiccation crack networks \cite{goehring2014cracking}. In temperate soils rich in clays and organic matter, repeated cycles and ongoing biophysical agitation produce robust millimeter-scale aggregates \cite{tisdall1982organic}, representing a natural form of granulation.  More generally, wetting–drying cycles draw heterogeneous particles into contact, while persistent mechanical and biological stirring prune unstable configurations, stabilizing isolated clusters that seed higher-order structure.\\

\noindent
\textsl{Memory Geomaterials.} The concepts of mechanical groups and environmental training together point toward a broader principle: geomaterials possess memory. In this context, memory geomaterials are Earth-mediated composites that retain the imprint of past physical, chemical, and biological conditions, which in turn shape their present and future mechanical responses; this encoded memory can be retained and retrieved. This idea aligns with the emerging notion of soil memory in earth system science, where soils are seen as dynamic recorders of past influences such as droughts, floods, wildfires, and land-use changes, as well as endogenous processes like microbial turnover and organic matter decomposition \cite{targulian2019soil,canarini2021ecological}. Memory in geomaterials is encoded at multiple levels. Abiotic carriers include mineral assemblages, pore structures, and residual organic matter, each reflecting past environmental perturbations. Biotic carriers such as microbial communities, worms, and higher fauna imprint memory through adaptation, compositional shifts, and functional diversification. Relic DNA, extracellular polymeric substances, and humus chemistry further provide molecular archives of past environmental conditions. These carriers operate across time (and length) scales ranging from hours (e.g., wet–dry cycles) to millennia (e.g., mineral transformations), embedding persistence and resilience into the geomaterial fabric \cite{canarini2021ecological,rahmati2023soil,li2019legacy}. 
 
The framework of materials geomimicry emphasizes that functional groups act as the primary drivers of microscale interactions, while environmental training prunes these interactions into robust macroscopic traits. Memory emerges when such traits preserve a record of prior environmental forcing. For instance, swelling versus non-swelling clay systems not only exhibit different elastic moduli but also retain signatures of hydration and structural reorganization that determine their long-term adaptation to stress. Similarly, repeated freeze–thaw or wet–dry cycles reorganize pore networks, establishing soil structures that persist and regulate hydrological and mechanical performance under new perturbations \cite{leuther2021impact,rooney2022soil}. The main concept is that environmental training, which happens through forcing, continuously shapes geomaterials in nature. Each episode of training leaves signatures embedded in the material’s structure. These persistent, structure-level imprints, that are the encoded record of past training, can be read out through mechanical protocols and are manifested in macroscopic response. Conceptually, memory is the intermediate imprint, which is written by the training protocol, stored in the hierarchical structure, and expressed as the final mechanical behavior. The recognition of memory in soft Earth materials has two central implications. First, it challenges the assumption that soil material properties can be inferred solely from their present composition: their response is conditioned by past exposures and evolutionary trajectories. Second, it suggests that laboratory models of soft matter, while useful, may fail to capture the inherent resilience and robustness exhibited by natural systems, which arise precisely because of memory effects. That begs the question, can we create such materials in the lab by complexifying simple model systems? 

\subsection*{Complexifying Model Soft Earth Systems}
In this section, we present a \emph{bottom-up} framework in which simple soft-matter systems are progressively ``complexified" and mechanically trained to demonstrate the emergence of Earth-mediated material properties. \\

\noindent
\textsl{Mixing Mechanical Functions.} Soil at the Earth’s surface is a soft, complex composite material composed of diverse fluid–particulate mixtures. Its fundamental building blocks -- sand, clay, silt, polymeric materials, microorganisms, and water -- combine in varying proportions to produce the final soil structure and composition, which in turn dictate macroscopic mechanics and transport functions. A zeroth-order model of soil can be engineered by mixing these basic functional groups to create the simplest complex system. For example, consider three essential ingredients from the list above: sand, clay, and water. These three building blocks introduce distinct mechanical functions: frictional, attractive, and viscous forces, respectively. Earlier work demonstrated that by varying the relative proportions of sand, kaolinite clay and water, one can produce tunable yielding behavior that helps to explain natural mud slides \cite{coussot1995structural,kostynick2022rheology,pradeep2024origins}. Furthermore, some of us showed for the first time in a material that brittle-to-ductile failure transitions can be programmed by progressively replacing the frictional component (sand) with the attractive component (clay) \cite{pradeep2024origins}. We showed that the full range of reported rheological behaviors for natural debris-flow materials could be reproduced by simply varying the proportions of our three-component mix. In dimensionless terms, the flow regime of these mixtures is governed by the viscous number $J = \dot{\gamma}\eta(\phi)/\tau_y$, where $\dot{\gamma}$ is the shear rate, $\eta(\phi)$ is the viscous dissipation as a function of particle concentration $\phi$, and $\tau_y$ is the suspension yield stress. The viscous number $J$ thus estimates the ratio of the grain rearrangement timescale to the macroscopic deformation timescale. These results demonstrate how a geomimicry approach can help to develop predictive models for natural hazards. 

Another case of complexifying soft soil-based matter is that of the most widely used material in the world, cement. Cement is produced by combining particulate phases with reactive chemistry; hydration products lock together frictional, attractive, and viscous interactions into a rigid, load-bearing network \cite{taylor1997cement,ioannidou2016mesoscale,Goyal2021physics}. This transformation from a tunable soft composite to a hardened solid illustrates how engineered complexity can create entirely new and robust mechanical functions \cite{ioannidou2016crucial}.\\

\noindent
\textsl{Tuning Environmental Stressors.} Creating complex materials by mixing functional groups produces macroscopic properties that differ from those of the individual building blocks. However, mechanical stresses in nature are often quite different from prescribe laboratory protocols. Thus, exploring lab-scale mechanical protocols that mimic environmental excitations (e.g., cyclic stressors) is crucial for understanding and tuning these simple mixtures as a class of adaptive learning systems. In this direction, we describe three distinct soil systems with increasing complexity in learning from mechanical stressing: creeping sandpile, solid bridges and mud cracks.

Recent tabletop experiments show that even an undisturbed sandpile on a flat surface undergoes slow, glass-like creep: its relaxation dynamics exhibit aging reminiscent of amorphous solids \cite{deshpande2021perpetual,hwang2016understanding,song2022microscopic}. Under continuous tapping, however, the pile evolves its microstructure in response to the imposed excitations. Brief episodes of tapping rapidly relax and strengthen the bulk (an annealing-like effect), while simultaneously exciting a thin, surface flowing layer that behaves like a landslide, where the bulk creep largely vanishes \cite{deshpande2021perpetual}. By contrast, cycling the same system through modest heating can rejuvenate the creep, restoring slow deformation \cite{deshpande2021perpetual,deshpande2024athermal}. More broadly, thermal cycling has been shown to systematically densify granular packings without mechanical input \cite{chen2006packing} and to induce creep motion in granular piles \cite{divoux2008creep}, establishing temperature fluctuations as an independent training protocol that complements mechanical forcing. Thus, even this minimal model of environmental forcing -- a sandpile on a table -- reveals rich learning behavior in which the dominant mechanical function is friction, yet the history of perturbations writes and rewrites the microstructural state that controls macroscopic response.

\begin{figure*}
\centering
\includegraphics[scale = 0.68]{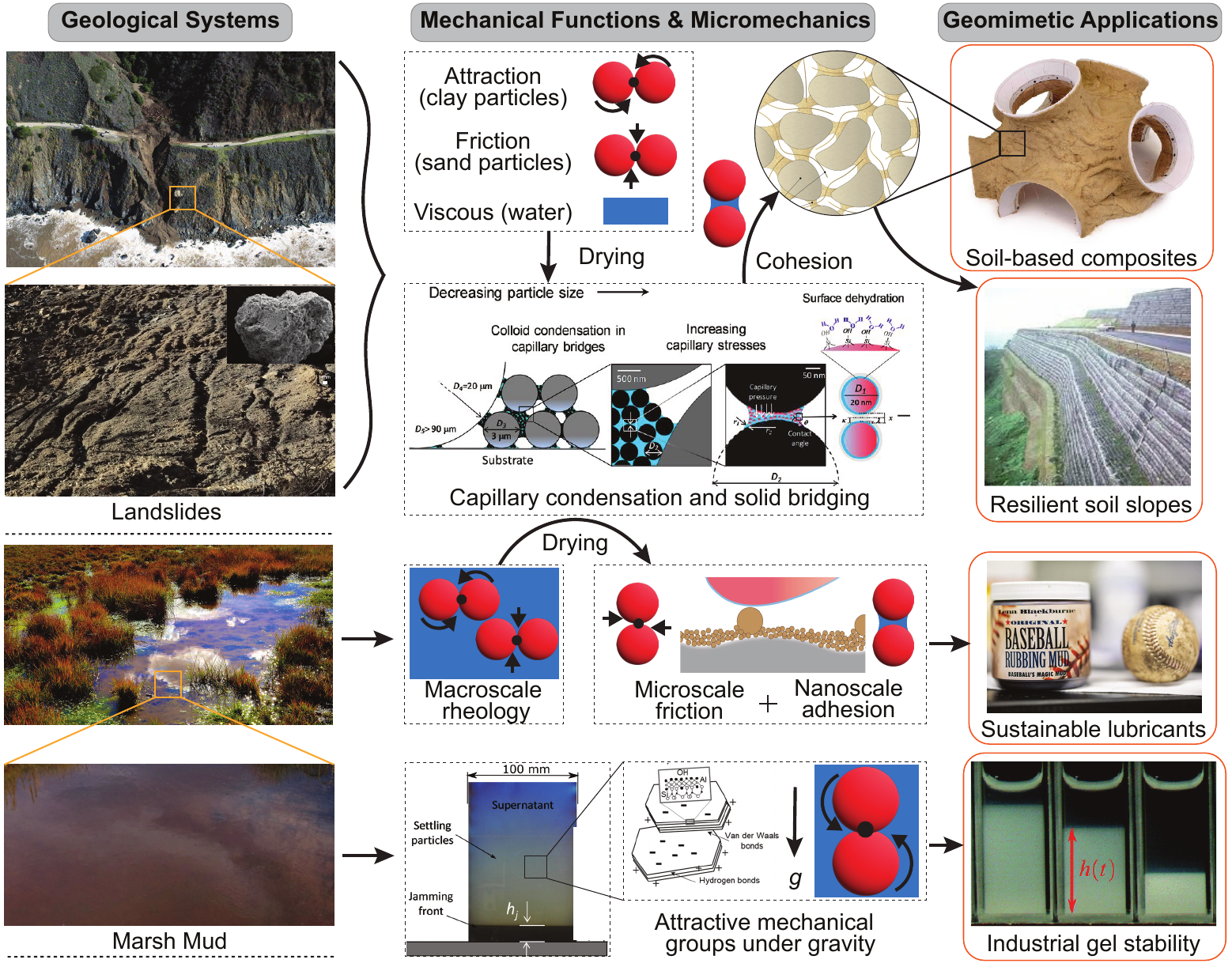}
\caption{\textbf{Mapping geological systems to geomimetic applications using mechanical functional groups.} We illustrate the complete geomimicry mapping using two natural systems explored in the literature: landslides and marshy soils. The image shows landslides in Big Sur, California (Courtesy: US Geological Survey). The flow and deformation of landslides can be modeled using a zeroth-order soil system composed of sand (frictional), clay (attractive), and water (viscous). Landslides and debris flows form rills that generate soil aggregates upon water evaporation (inset SEM \cite{brevik2015interdisciplinary}). Similarly, the model soil system dries to produce aggregates resembling those found in natural rills \cite{kostynick2022rheology}. The mechanism of resilient aggregate formation follows fractal-like capillary condensation, creating solid bridges across multiple length scales \cite{seiphoori2020formation}. Here, a new mechanical function -- cohesion -- emerges, which can be leveraged to engineer resilient, soil-based composites with tunable mechanical properties by adjusting the attraction-to-friction ratio and modulating drying dynamics \cite{lasting2024terrene} (Courtesy: Federal Highway Administration). Marshy soils, which contain a high concentration of clay (attractive component), exhibit strong shear-thinning behavior and can be easily processed. Upon drying, sparse sand particles impart microscale friction, while the clay matrix provides nanoscale adhesion. This combination of flow, friction, and adhesion properties makes marshy muds promising for sustainable lubricant applications \cite{pradeep2024soft}. Moreover, marshy muds exist as fragile gels, which can be studied in the laboratory to examine how mechanical functional groups evolve under gravitational settling \cite{seiphoori2021tuning}. Insights from such studies can inform the understanding and stabilization of pharmaceutical formulations and other industrial products \cite{harich2016gravitational}.
}
\label{fig4}
\end{figure*}

Experiments simulating wet–dry cycles on model mono- and poly-disperse clay particles reveal emergent system features that cannot be predicted from the dry mixtures, or from conventional rheological measurements: (i) evaporation drives condensation of smaller particles within shrinking capillary bridges among the larger grains, with the suction pressure eventually forming stabilizing “solid bridges” of smaller particles that bond larger particles together. This is a process that repeats across length scales down to the van der Waals well of the smallest particles; and (ii) the discovery of a characteristic cohesion length scale ($\sim$5 $\mu$m) and that particle size, rather than intrinsic material properties, governs the cohesive strength of aggregates formed by evaporation \cite{seiphoori2020formation}. These results suggest a minimal model for the formation of natural soil aggregates; intricate, hierarchical structures can be assembled from a polydisperse suspension by simply letting it evaporate. The characteristic cohesion length scale ($\sim 5 \mu m$) identified in these experiments can be understood through a granular Bond number, $Bo_g = F_{cohesive}/F_{gravity}$, which compares interparticle adhesive forces to the weight of a grain. For particles smaller than $\sim 5 \mu m$, $Bo_g >> 1$ cohesion dominates, enabling stable bridge formation; above this size, $Bo_g << 1$ and gravity disrupts cohesive contacts. This crossover sets the fundamental building-block size for soil aggregate assembly.  Our findings help to explain the emergence of cohesion and structure in soils, and may lead to new strategies for stabilizing soils to resist erosion \cite{seiphoori2020formation}. At the same time, this geomimicry approach may also help in predicting the role of moisture in powder handling across industrial applications.

Another example where environmental stressors “train” soft matter is the evolution of mud cracks under repeated wet–dry cycles. Similar to above, a model mud can be constructed with three ingredients: sand, a swelling clay like bentonite, and water. Initial drying of this model mud creates rectilinear, T-junction–dominated networks as shrinking drives crack formation. Repeated wetting and drying causes this network to reorganize, as each cycle of swelling and shrinking shifts crack positions; T junctions ``twist'' into Y junctions over several generations \cite{goehring2010evolution}. This learning-by-cycling is local and history-dependent; prior cracks guide new ones, orders of opening change, and vertices drift in the direction of the formerly orthogonal branch. The accumulated drift scales with layer thickness, providing a geometric memory of environmental forcing \cite{goehring2013evolving,goehring2010evolution,goehring2014cracking,ma2019universal}. Occasionally cracks even “forget” and reappear after several cycles, underscoring a persistent yet plastic memory \cite{goehring2010evolution}. Despite this apparent complexity, there is a general trend: mudcracks subject to repeated wetting and drying evolve toward a configuration that minimizes deformation under cyclic strain; this configuration is Voronoi tesselation \cite{silver2025decoding}. A simple dynamical model, that codifies the emergent dynamics of cracking and twisting with simple rules, is capable of reproducing the geometric properties of natural mudcracks \cite{silver2025decoding}. 

Beyond wet–dry cycling, systematic multiscale exploration of subtler or coupled environmental excitations, such as freeze-dry, small humidity fluctuations, thermal cycles, earthquake-like cyclic loading etc., remains largely unexplored and may reveal new routes to program learning and mechanics in complex soft earth systems. \\

\noindent
\textsl{Imprinted Material Memory.} Soil systems evolve to acquire properties that reflect their structure and composition, shaped by the dynamic influence of environmental stressors; soils can evolve acquired properties as their structure and composition reorganize under environmental stressors. To understand how such “memory” is encoded and read out, we turn to amorphous and jammed model systems under cyclic shear. Near yield, disordered solids self-organize into reversible plastic limit cycles: particles undergo local, hysteretic rearrangements each half-cycle that dissipate energy yet return stroboscopically to their original configuration, thus providing a clear microscopic signature of (local) stored mechanical memory distinct from global irreversibility. This separates microstructural yielding (onset of irreversible change) from rheological yielding (growth of dissipation), and shows that many plastic events can encode history without permanent flow \cite{keim2013yielding,keim2014mechanical,galloway2020quantification,galloway2022relationships}. 

Such memories include directional memories linked to the organization and orientation of shear-transformation–like regions that repeatedly rearrange under cycling. Thermal fluctuations weaken these memories: even minimal Brownian noise destabilizes reversible cycles and increases transitions to irreversibility, highlighting that memory strength depends on both drive amplitude and noise level (and temperature). Together, these results establish a practical toolkit for writing memory (controlled cyclic loading), reading it (tracking non-affine, hysteretic rearrangements), and erasing it (over-cycling well above yield or raising noise) \cite{farhadi2017shear,galloway2020quantification,galloway2022relationships}. Even simple shear flow, however, can help to build memory in these yield stress fluids, leading to complex relaxation and residual stresses which carry the information of the flow history [184]. Such relaxation can take the form of non-monotonic temporal evolutions, as demonstrated in colloidal gels where the stress first decays and then increases after flow cessation before settling to a history-dependent residual value \cite{sudreau2022residual}, a phenomenon rationalized through the differential relaxation of shear-banded states \cite{ward2025shear}.

The imprints of memory in disordered materials (jammed particulate suspensions) can also be observed at the structural level in terms of local packing which adapted to the flow or other history \cite{vinutha2024memory,vasisht2020emergence}. Using the concept of excess entropy as a structural parameter to describe the level of caging in athermal particle suspensions,  it was recently found the material ``remembers" and faithfully follows the external forcing in a sinusoidally-driven interfacial stress rheometer if deformations are elastic (i.e., below yielding) \cite{galloway2022relationships}. As the strain amplitude is increased above yielding, the structural parameter (i.e., excess entropy) signal deviates from the forcing signal, indicating that the structure is ``forgetting" its sinusoidal imprint. More precisely, memory erasure is not simply a consequence of exceeding the yield point. It depends on the ratio of shear forces to interparticle attraction, captured by the Mason number, \emph{Mn}, as well a micro-structural configurational changes \cite{varga2018large,varga2019hydrodynamics,sudreau2022shear,das2021shear}. Complete rejuvenation requires fully fluidizing the sample ($Mn >> 1$), whereas partial structural memory may persist at intermediate values of \emph{Mn} near yielding. These results provide a quantitative bridge from microstructure to dissipation, i.e. from how memory is stored to how it shapes bulk rheology, and motivates modeling yield and brittle-ductile response directly from microstructural metrics. Such a framework suggests clear routes to translate these memory principles to soft earth systems: cyclic environmental forcing can write directional memories in soil aggregates, while microstructural descriptors (e.g., pair correlations related to excess entropy) can serve as experimentally accessible readouts of the learned state \cite{galloway2020scaling,galloway2022relationships}.

Soft Earth analog laboratory systems likewise display memory effects. In particular, we highlight recent findings from studies of fluid-sheared granular beds composed of index–matched particles. Under sustained subcritical forcing, beds undergo strain hardening: creep compacts the packing and imprints an anisotropic fabric, with both mechanisms contributing comparably to increased resistance \cite{cunez2022strain}. This structural evolution raises the threshold fluid stress for entrainment, expressing a stored memory of prior loading. Conversely, episodic fluidization acts as an erasure protocol, softening the bed and resetting the threshold. These findings provide a mechanistic basis for field observations that entrainment thresholds in natural rivers increase following prolonged subcritical flows \cite{masteller2025modeling}.

Emergence of memory in complex soft-earth–like materials is a non-trivial phenomenon. Beyond largely qualitative field observations, systematic studies of mechanically encoded memory in particulate media have been conducted predominantly in idealized systems. These are typically near monodisperse, spherical particles under controlled protocols. Extrapolating from these models is dangerous: compositional and geometric heterogeneity could enhance memory by promoting hierarchical organization and multiscale load paths, or it could erase memory by introducing disorder that obscures reproducible structural states. Discriminating between these regimes requires targeted experiments that vary heterogeneity in a controlled manner while jointly resolving structure across scales and macroscopic response.

\section*{Geomimicry Design Rules}

We cast geomimicry as a training problem, where the microstructural operators are programmed from external environmental forcings that are eventually encoded as ``traits" of Earth-mediated matter. In the \emph{top-down} view introduced earlier, exposures such as humidity, thermal cycling, flooding, and wind act as training methodologies that select among many admissible microstructures to create final soil microstructure. In the complementary \emph{bottom-up} view, we deliberately assemble mechanical functional groups and then train the mixture so that its final properties \textit{emerge}, which has the embedded signatures of perturbations, rather than only its starting composition. The practical objective is to formulate design rules that map: (i) a specified distribution of environmental stresses; (ii) controllable reconfiguration pathways of constituent interactions; and (iii) measurable macroscopic functions (flow and deformation, failure modes, frictional dissipation, transport dynamics etc.) that reads out these embedded signatures of forcings. This framing completes geomimicry: choose the environmental pressure protocol to write the trait. In particular, the \emph{top-down} asks the following questions: what pressures shaped the material? how did forces reconfigure microstructure?, and what functions emerged? Answers to these queries become design levers to encode mechanical functions in soft geomaterials through environmental training techniques. We are, however, still in the early stages of developing such design rules, and only guidance is provided here. 

As a starting point one needs to identify environmental pressures. Directed aging and cyclic conditioning provide complementary routes to encode traits in Earth-mediated mixtures. Directed aging uses biased stresses (e.g. sustained compression or humidity gradients) to tilt the energy landscape so that the system relaxes quickly into configurations that produce a desired response, while cyclic conditioning near yield writes reversible, hysteretic rearrangements that act as local “memory bits” \cite{pashine2019directed}. These amorphous soft earth materials are effectively topologically disordered networks, and training acts by evolving structural configurations across scales, thus enabling target network responses (e.g. breaking and forming force chains, cohesive hydroclusters etc.) rather than the macroscopic continuum. In networked solids this has been used to imprint unusual elastic functions \cite{shivers2025criticality,shivers2020compression,reid2018auxetic} and should translate to pore-scale architectures in soils. Finally, macroscopic functions emerge from the training and microstructure. These protocols integrate naturally with the bottom-up mixing of functional groups: by sweeping solid volume fraction ($\phi$) and clay fraction ($\chi$) to set the baseline frictional/attractive/viscous balance, then applying tailored cycles (amplitude, frequency, sequence), one can program transitions such as brittle-to-ductile failure or tune yield stresses in soil analogs \cite{pradeep2024origins}. The protocol, in addition to the soil composition, selects the final ``trait". 

A natural question is to what extent laboratory scale experiments can be considered analogous to geological processes that occur over millennia. We argue that the relevant quantity here is the Deborah number, defined as $De = t_{relax} / t_{forcing}$. This estimates the material's structural relaxation timescale to the period of environmental forcing. When $De$ and other relevant dimensionless groups are matched between the lab and field, we posit that the cumulative effect of cyclic forcing on the microstructural evolution can be considered mechanistically equivalent. This view is consistent with the ``unreasonable effectiveness" of stratigraphic and geomorphic experiments demonstrated by Paola et al. \cite{paola2009unreasonable}, who argued that the laboratory systems reproduce geological dynamics despite the orders-of-magnitude differences in absolute timescales. This is due to the emergent, self-organized dynamics of Earth systems being naturally scale-independent. Laboratory scale experiments should therefore be understood not as a compressed version of geological processes, but as a means of accessing the same dynamics that natural systems exhibit over longer timescales.

More generally, the phenomena described in this perspective are governed by a set of dimensionless numbers that quantify the competition between relevant forces at different scales. As discussed earlier, the Shields number ($\theta$) governs whether fluid stresses can mobilize grains against gravity, the viscous number ($J$) distinguishes soft soil flow regimes, the granular Bond number ($Bo_g$) Determines whether cohesive or gravitational forces dominate at the particle scale, the Mason number ($Mn$) sets whether shear can overcome interparticle attraction to erase structural memory, and the Deborah number ($De$) controls whether memory persists or relaxes between forcing events. In addition, shear P\'eclet number ($Pe = \dot{\gamma}R^2/D_0$) comparing advective transport of particle of average radius $R$ to Brownian diffusion ($D_0$ is the diffusion coefficient as a function of $\phi$), becomes relevant for the colloidal-scale constituents (e.g., clays and fine silt) in these mixtures. Environmental training protocols act by driving these dimensionless groups through critical thresholds. A quantitative geomimicry framework will ultimately require mapping how these dimensionless groups co-evolve under realistic, broadband environmental forcing.

In summary, adaptability-based design rules for geomimicry treat the material platform, structural disorder or network priors, training protocols, readout metrics, and retraining schemes as equally important. This framework transforms environmental history from a source of uncontrolled variability into a deliberate driver of emergent functions.

\section*{Applications}

\begin{figure*}
        \centering
        \includegraphics[scale = 0.75]{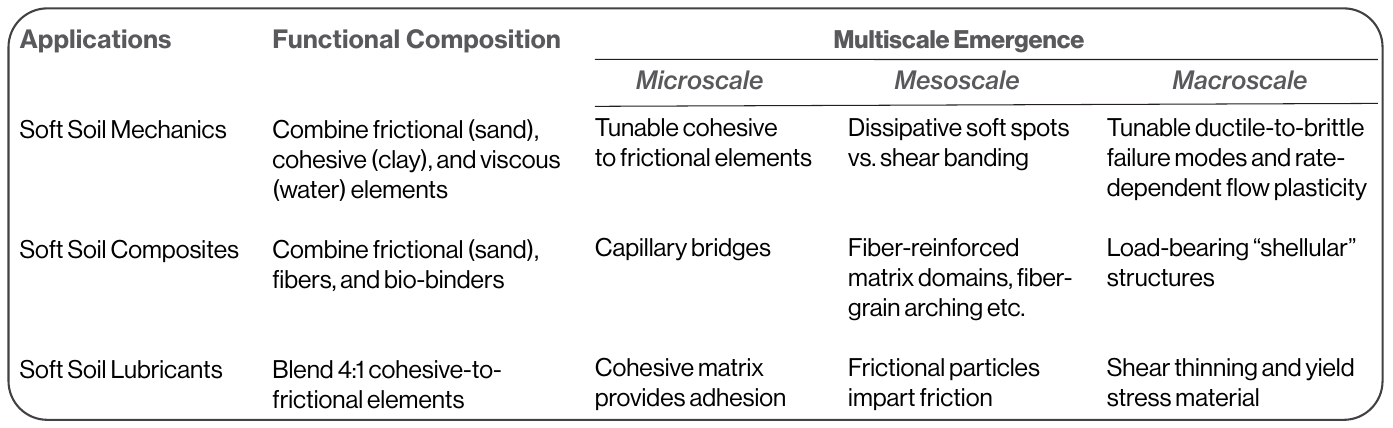}
        \caption{\textbf{Multiscale emergence of mechanical properties in geomimetic applications.} The geomimicry framework links functional composition to emergent properties across micro-, meso-, and macroscales. In soft soil mechanics, combining frictional, cohesive, and viscous elements creates tunable microscale interactions that dictate mesoscale dissipative soft spots or shear banding, programming macroscale ductile-to-brittle failure modes and rate-dependent flow plasticity. For soft soil composites, integrating sand, fibers, and bio-binders drives microscale capillary bridges and mesoscale fiber-reinforced domains to form macroscale load-bearing "shellular" structures. In soft soil lubricants, cohesive-to-frictional blend utilizes a microscale cohesive matrix and mesoscale frictional particles to produce a macroscale shear-thinning, yield-stress material.}
        \label{fig5}
    \end{figure*}

Here, we apply materials geomimicry to Earth-based systems. By tuning constituent materials, we can engineer adaptive composites and decouple the multiscale emergence of properties in natural soils (Fig. \ref{fig5}).\\

\noindent
\textsl{Soft soil mechanics.} Framing geomechanical challenges as soft-matter problems yields fundamental insights into soil mechanics. A water-saturated sand-clay mixture provides a minimal three-phase functional blend of frictional, attractive, and viscous elements \cite{pradeep2024origins}. Under shear, the competing rearrangement timescales of the cohesive and frictional elements govern macroscopic failure. By tuning the relative volume fractions of these components, one can program flow plasticity and select ductile-to-brittle failure modes. This mechanical function-controlled framework resolves prior rheological disagreements by cleanly distinguishing the failure mechanisms of sand-rich versus clay-rich debris flows \cite{kostynick2022rheology,coussot1995structural}.\\

\noindent
\textsl{Soil-based composites.} The mechanical properties of bio-derived soil composites can be manipulated by combining distinct mechanical functions and tuning processing conditions. First, mixing frictional sand with reinforcing fibers and bio-binders modulates the frictional response, enabling 3D printing and extrusion. Second, post-printing drying activates an attractive mechanical function as capillary bridges and binder curing develop inter-particle cohesion. The resulting composite performance is intrinsically multiscale: microscale capillary bridges enhance cohesion, mesoscale fiber networks suppress crack propagation, and macroscale ``shellular" geometries distribute stresses efficiently \cite{Gao2003materials,Chawla2025_RAMUS,monfared2020effect,hejazi2012simple}. Collectively, this transforms granular matter into an adaptive structural material with tunable failure modes.\\

\noindent
\textsl{Soil-based lubricants.} Natural soils can modify surface texture to tune processability and haptic response, as demonstrated by Major League Baseball's rubbing mud. This specific blend relies on a cohesive-to-frictional functional ratio of approximately 4:1 by volume. The dominant cohesive function confers shear-thinning during application, promoting smooth spreading and uniform coverage. Upon drying, the sparse frictional phase embedded within the cohesive matrix elevates mesoscale dynamic friction by $\approx$50\% and increases nanoscale stickiness by $>$100\% \cite{pradeep2024soft}. This coordinated interplay of cohesive, frictional, and viscous elements yields an application-specific balance of spreadability and grip.

The above are just a few examples on how the geomimicry framework can lead to novel, sustainable materials with exquisite functionalities. There are, of course, other examples in the literature and already in industry. But the point is to be deliberate in the approach. 

\section*{Discussion and Implications}

Framing complex soil materials in terms of mechanical functions such as cohesion, friction, squishiness (compressible compliance, elasticity), and viscosity, provides a common language for emergence and universality across systems that differ in chemistry and microstructure. It is important to delineate the scope of the geomimicry framework. The paradigm is most naturally applicable to amorphous, multicomponent geomaterials that possess accessible configurational degrees of freedom under environmental forcing. That is, systems whose microstructure can be reorganized without catastrophic failure. This includes soils, unconsolidated sediments, fault gouges, shales (in certain cases), and planetary regolith. Weakly cemented or poorly lithified rocks retain some capacity for training. The framework is less applicable to intact crystalline rocks, where mechanical behavior is governed by crystal structure and intragranular dislocation mechanics, or to sedimentary rocks, whose microstructure has been locked into rigid configurations. In general, geomimicry applies where the physics is dominated by soft particulate mechanics, such as contact mechanics, capillary forces, and colloidal scale interactions, that introduces particle-scale rearrangement timescales and where the material retains sufficient structural disorder to encode memory of environmental forcings. This abstraction enables model-analogous platforms: carefully mixed soft-matter constituents that implement the same functional set can replicate the salient mechanics. In turn, these analogs let us decouple and tune interactions, interrogate structure–property links across scales, and export design rules back to natural and engineered soils.

In soft-soil engineering, the framework suggests practical routes for conditioning soils against erosion, improving constructability, and modulating surface texture. A particularly tunable lever is solid bridging, wherein small inert particles, coupled with controlled wetting, form capillary or binder-mediated links that elevate cohesive function without sacrificing processability. This principle directly informs climate-resilient surface treatments, powder-lubricant formulations, and texture modulators. The same soft-soil mechanics, cast in mechanical functional terms, extends to multicomponent soft matter central to sustainability and decarbonization \cite{Habert2020environmental}. Two important classes of materials are meat-alternative food matrices and lithium-battery electrolytes (slurries and gels), where rheology governs processing (mixing, extrusion, coating) and tribology governs performance (mouthfeel, separator/electrode interaction). Designing functional blends that balance cohesion (network strength), friction (particulate contact), and viscosity (carrier phase) can rationalize processing windows and end-use properties, while offering tunability for waste reduction and energy efficiency.

Coupling soft-soil mechanics with solid bridging has far-reaching implications for precision agriculture. For example, understanding how water modifies different types of soil can be understood in terms of mechanical functional groups and inform humidity sensors about the quality of data obtained. The same principles translate to wheel and leg interactions in off-road and planetary exploration, where adaptive traction, reduced wear, and tunable compaction thresholds can be engineered by writing specific functional memories (via wetting-drying, vibration, or temperature cycles) into the regolith analog \cite{bush2024relating,ruck2025unified}. Concepts from mud-crack evolution point to programmable cracking in thin films and nanostructured coatings: by prescribing drying kinetics, prestress, and substrate adhesion, one can encode crack geometries that serve as functional patterns (e.g., microfluidic channels, optical textures) \cite{ma2022programming}. Conversely, geometric statistics of crack networks may act as environmental proxies, enabling inference of climate histories on planetary surfaces \cite{silver2025decoding}. 

Together, understanding soil-based materials from a mechanical functional group perspective does more than unify descriptions. By selecting mechanisms to implement targeted mechanical functions, and by training those functions through environmental protocols, we obtain a compact, transferable framework for engineering soft soils and their analogs across soil on the Earth, planetary bodies, and industrial applications.

\section*{Acknowledgements}
The authors are grateful for the support from the US National Science Foundation Division of Materials Research (NSF-DMR-2422537) to P.E.A and D.J.J, NSF Engineering Research Center for the Internet of Things for Precision Agriculture (NSF-EEC-1941529) to P.E.A, NSF-DMREF (CBET-2118962) and NSF-DMR (Grant 2226485) to E.D.G, NASA Planetary Science and Technology Through Analog Research Program (PSTAR-80NSSC22K1313) and NASA Lunar Surface Technology Research (LuSTR-80NSSC24K0127) to D.J.J., and University of Pennsylvania Center for Soft and Living Matter Postdoc Fellowship to S.P.

\bibliographystyle{unsrt}
\bibliography{References}

\end{document}